\documentclass[aps,pra,reprint,superscriptaddress]{revtex4-1}

\usepackage{amsmath}    
\usepackage{graphicx}   
\usepackage[lofdepth,lotdepth, caption=false]{subfig}
\usepackage{verbatim}   
\usepackage{color}      
\usepackage{hyperref}   
\usepackage{dcolumn}
\usepackage{bm}
\usepackage{miller}

\begin{document}
\newcommand{\degrees}{$^\circ$C}

\title{Phonon coupling to dynamic short-range polar order in a relaxor ferroelectric near the morphotropic phase boundary}
\author{John A.~Schneeloch}
\email{jschneeloch@bnl.gov}
\affiliation{Condensed Matter Physics and Materials Science Department, Brookhaven National Laboratory, Upton, New York 11973, USA}
\affiliation{Department of Physics and Astronomy, Stony Brook University, Stony Brook, New York 11794, USA}

\author{Zhijun Xu}
\affiliation{Physics Department, University of California, Berkeley, California 94720, USA}
\affiliation{Materials Science Division, Lawrence Berkeley National Laboratory, Berkeley, California 94720, USA}

\author{B.~Winn}
\affiliation{Quantum Condensed Matter Division, Oak Ridge National Laboratory, Oak Ridge, Tennessee 37831, USA}

\author{C.~Stock}
\affiliation{School of Physics and Astronomy, University of Edinburgh, Edinburgh EH9 3JZ, United Kingdom}

\author{P.~M.~Gehring}
\affiliation{NIST Center for Neutron Research, National Institute of Standards and Technology, Gaithersburg, Maryland 20899, USA}

\author{R.~J.~Birgeneau}
\affiliation{Physics Department, University of California, Berkeley, California 94720, USA}
\affiliation{Materials Science Division, Lawrence Berkeley National Laboratory, Berkeley, California 94720, USA}

\author{Guangyong Xu}
\affiliation{Condensed Matter Physics and Materials Science Department, Brookhaven National Laboratory, Upton, New York 11973, USA}
\date{\today}

\begin{abstract}
We report neutron inelastic scattering experiments on single crystal PbMg$_{1/3}$Nb$_{2/3}$O$_{3}$ doped with 32\% PbTiO$_{3}$, a relaxor ferroelectric that lies close to the morphotropic phase boundary. When cooled under an electric field $\mathbf{E} \parallel$ [001] into tetragonal and monoclinic phases, the scattering cross section from transverse acoustic (TA) phonons polarized parallel to $\mathbf{E}$ weakens and shifts to higher energy relative to that under zero-field-cooled conditions.  Likewise, the scattering cross section from transverse optic (TO) phonons polarized parallel to $\mathbf{E}$ weakens for energy transfers $4 \leq \hbar \omega \leq 9$ meV.  However, TA and TO phonons polarized perpendicular to $\mathbf{E}$ show no change.  This anisotropic field response is similar to that of the diffuse scattering cross section, which, as previously reported, is suppressed when polarized parallel to $\mathbf{E}$, but not when polarized perpendicular to $\mathbf{E}$. Our findings suggest that the lattice dynamics and dynamic short-range polar correlations that give rise to the diffuse scattering are coupled.

\end{abstract}

\maketitle


\section{Introduction}
Relaxor ferroelectrics have great potential for applications due to their piezoelectric and dielectric properties \cite{park_ultrahigh_1997, uchino_piezoelectric_1996, service_shape-changing_1997}, but there is much that is not understood about how their properties arise on a microscopic level \cite{xu_competing_2010}. The composition of many relaxors is related to simple perovskites with the ABO$_{3}$ formula, but with one cation site randomly filled with two or more cations of different valences, resulting in strong, disordered electric fields. For the relaxor PbMg$_{1/3}$Nb$_{2/3}$O$_{3}$ (PMN), the B$^{4+}$ site is occupied by Mg$^{2+}$ and Nb$^{5+}$. The variation in valence can be reduced by doping with an ion of intermediate valence, e.g., Ti$^{4+}$ in the case of PMN doped with $x$\% PbTiO$_{3}$ (PMN-x\%PT). The resulting PMN-$x$\%PT phase diagram shows four basic regions \cite{kiat_monoclinic_2002, cox_universal_2001, noheda_phase_2002}: a cubic paraelectric phase at high temperature for all $x$; a region with relaxor behavior for low $x$ with either cubic or rhombohedral symmetry; a tetragonal, conventional ferroelectric region for high $x$; and a morphotropic phase boundary (MPB) region between the relaxor and tetragonal regions. The piezoelectric coefficients $d_{33}$ are very large in the MPB region and abruptly drop for higher $x$ \cite{guo_phase_2003, park_ultrahigh_1997, kuwata_phase_1981, kuwata_dielectric_1982}; understanding this behavior and exploiting the large piezoelectricity provide much of the motivation for exploring relaxor ferroelectrics. These PMN-$x$\%PT solid solutions with small $x$ exhibit clear relaxor behavior characterized by large dielectric constants which have a broad maximum with respect to temperature and are highly frequency-dispersive within this range. These relaxor behaviors are widely believed to be associated with polar nano-regions (PNR) or other short-range polar order, as shown by numerous x-ray and neutron diffuse scattering studies \cite{xu_probing_2011, hiraka_cold_2004, you_diffuse_1997, takesue_x-ray_2001, xu_three-dimensional_2004, la-orauttapong_diffuse_2001, la-orauttapong_neutron_2003, hirota_neutron_2002, dkhil_local_2001, vakhrushev_unknown-title_1995, hlinka_diffuse_2003, vakhrushev_diffuse_2005, xu_neutron_2004}.

We have previously characterized two distinct components of the diffuse scattering in PMN-$x$\%PT, which we label T1 and T2 \cite{xu_two-component_2010} as shown in Fig.\ \ref{fig:fig1}(a). These labels are intended to refer to the related phonons where a T1 mode is a transversely polarized phonon propagating along $\langle 100 \rangle$ and a T2 mode is a transversely polarized phonon propagating along $\langle 110 \rangle$. For example, near (100), a TA$_{1}$ phonon mode would refer to the transverse acoustic phonon mode propagating along the [010] or [001] directions with polarization along [100], while near (110), a TA$_{2}$ phonon mode would refer to the transverse acoustic phonon mode propagating along [1$\bar{1}$0] with [110] polarization. The two diffuse scattering components can be distinguished by their anisotropic response when field-cooled (FC), i.e. after applying electric field above the ferroelectric transition temperature T$_{c}$ and then cooling below T$_{c}$. With a field applied along [111], a redistribution of T2-diffuse scattering intensity between two differently-oriented components polarized along [110] and [1$\bar{1}$0] has been observed in the structurally similar perovskite PbZn$_{1/3}$Nb$_{2/3}$O$_{3}$ doped with $x$\% PbTiO$_{3}$ (PZN-$x$\%PT) \cite{xu_persistence_2005, xu_coexistence_2006, xu_electric-field-induced_2006} and in PMN \cite{stock_neutron_2007}. This situation has been interpreted in terms of a domain effect, in which applying a [111] field creates a single [111]-polarized ferroelectric domain (as opposed to the eight possible $\langle$111$\rangle$-polarized domains present in the zero-field cooled state) which favors certain orientations of polar nanoregions (PNR), resulting in the redistribution of diffuse scattering intensities along certain $\langle$110$\rangle$ directions \cite{xu_electric-field-induced_2006}. An electric field along [001], on the other hand, does not seem to significantly affect the T2-diffuse scattering in the $H0L$ plane \cite{wen_response_2008}. 

Transverse acoustic phonons propagating along $\langle 110 \rangle$ (TA$_2$-phonons) are expected to couple with T2-diffuse scattering modes \cite{stock_strong_2005}. Evidence for this diffuse-TA$_2$ phonon coupling has been shown with the help of an external [111] electric field, which breaks the pseudo-cubic symmetry and reveals a clear difference between TA$_2$ phonons measured near (220) and (2$\bar{2}$0)~\cite{xu_phase_2008}. T1-diffuse scattering, on the other hand, does not show a redistribution of scattering intensity under an external field. Instead, a suppression of [001]-polarized T1-diffuse scattering occurs under [001]-field cooling, while the [100]-polarized T1-diffuse scattering remains unaffected, as has been shown in PZN-$x$\%PT \cite{gehring_electric-field_2004, xu_two-component_2010}. Coupling between the T1-diffuse modes and the TA$_1$ phonon modes has not yet been thoroughly studied.

In this paper we report neutron scattering experiments on PMN-32\%PT with a field applied along [001]. When cooled below $T_{c} \approx 430$ K~\cite{wen_response_2008}, in addition to the expected suppression of T1-diffuse scattering measured near (001), we also see a clear change in the intensities of the TA$_1$ phonons near (001), whereas the TA$_{1}$ phonons near (100) are unaffected. These changes illustrate that there may be a TA$_1$-phonon/T1-diffuse mode coupling, evocative of the TA$_2$-phonon/T2-diffuse mode coupling previously seen in PZN-4.5\%PT \cite{xu_phase_2008}. This coupling appears to be limited to large-wavelength phonons. In addition, we observed a suppression of spectral weight for the transverse optic phonons in the T1 direction (TO$_1$ phonons) within 4 to 9 meV near (002), but no change was seen near (200). The changes in the TA$_1$ and TO$_1$ phonons were present at 400 K but much less pronounced at 200 K.

\section{Experimental Details}
We purchased a PMN-32\%PT single crystal from TRS Ceramics with dimensions $10 \times 10 \times 2$ mm$^{3}$ and large [001] faces. The (001) surfaces were coated with gold to ensure a uniform equipotential surface during field application. Another PMN-32\%PT crystal from the same source was measured to have a cubic-tetragonal transition at $T_{c} \approx 430$ K and a tetragonal-monoclinic transition near 355 K\cite{wen_response_2008}. Neutron inelastic scattering experiments were performed on the HYSPEC time-of-flight spectrometer at the Spallation Neutron Source at Oak Ridge National Laboratory \cite{stone_comparison_2014}. The software package Mantid was used in the processing of the data \cite{arnold_mantiddata_2014}. The incident energy $E_{i}$ was set to 20 meV. The crystal symmetry was pseudocubic with lattice parameter $a = 4.00$ \AA. All neutron scattering momentum transfers $\mathbf{Q}$ are reported in terms of reciprocal lattice units (r.l.u.), and energy transfers $\hbar \omega$ are reported in meV. Measurements were performed in the $H0L$ scattering plane. Fields of 0.5-8.0 kV/cm were used. The T1-diffuse scattering at (001) changed significantly with a field of 0.5 kV/cm, and had almost no additional change with higher field, indicating that 0.5 kV/cm was sufficient to alter the T1-polarized short-range order. For pseudocolor plots, the data were smoothed. Errorbars represent statistical error and correspond to 1 standard deviation from the observed value.


\section{Data and Analysis}

\begin{figure}[h]
\begin{center}
\includegraphics[width=8.6cm]
{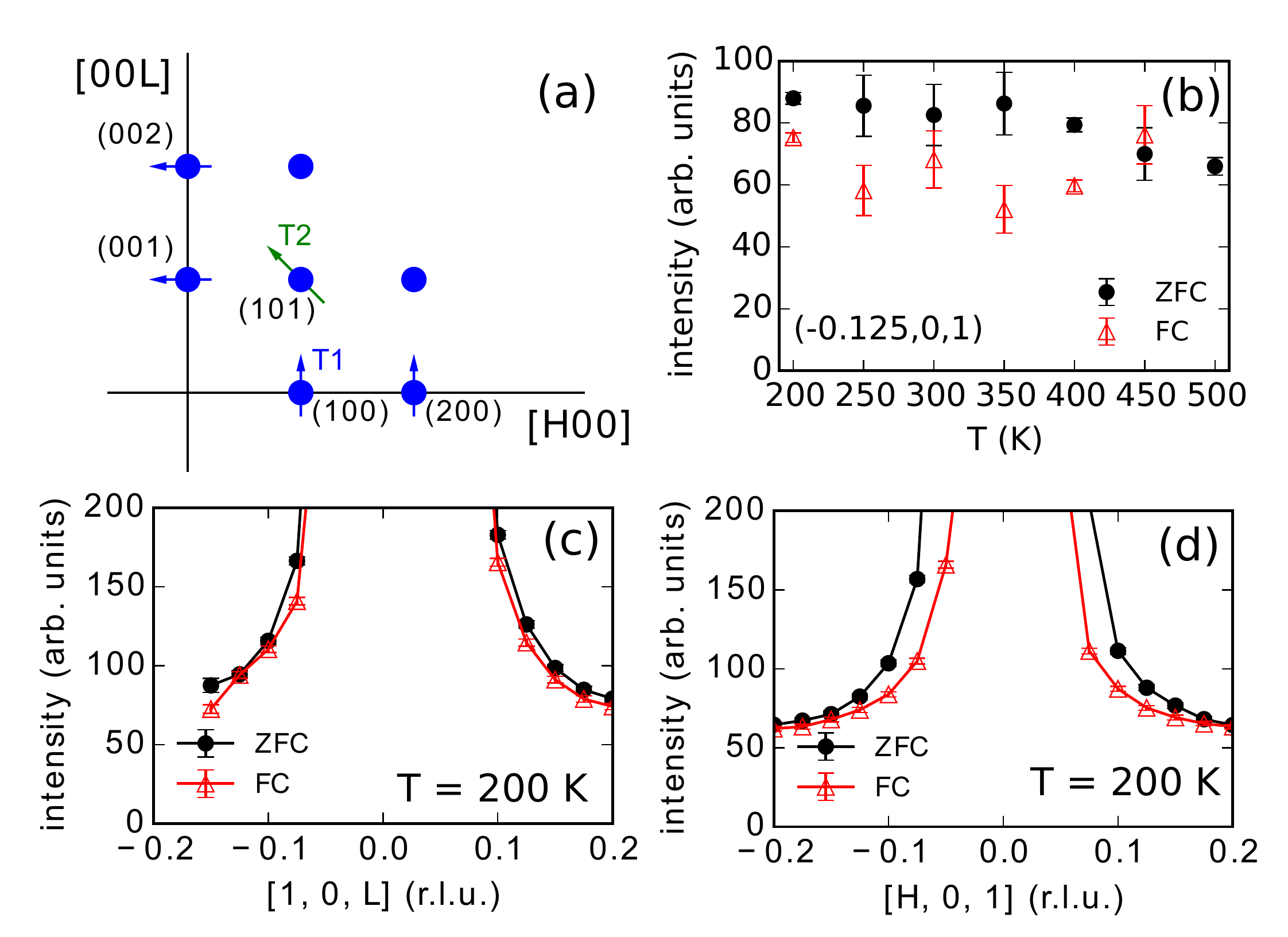}
\caption{\label{fig:fig1} (Color online.) (a) Schematic diagram of the $H0L$ plane in reciprocal space. The blue and green arrows indicate the T1 and T2 directions, respectively. (b) Temperature dependence of T1-diffuse neutron scattering, shown via intensity at wavevector $\mathbf{Q} = (-0.125,0,1)$ and energy transfer $\hbar \omega$ = $0$ meV plotted as a function of temperature. (c,d) Field-dependence of diffuse scattering near (100) and (001), shown from elastic neutron scattering intensity plotted along [10$L$] and [$H$01] at 200 K for zero-field-cooled (ZFC) and field-cooled (FC) conditions.}
\end{center}
\end{figure}

In Fig.\ \ref{fig:fig1}(b), the temperature dependence of the T1-diffuse scattering near (001) is shown with neutron scattering intensity measured at wavevector $\mathbf{Q} = (-0.125, 0, 1)$ for FC and zero-field-cooled (ZFC) conditions. The intensity was integrated within $0.95 \leq L \leq 1.05$ r.l.u., $-0.1375 \leq H \leq -0.1125$ r.l.u., and $-0.5 \leq \hbar \omega \leq 0.5$ meV. For temperatures up through 400 K a clear suppression of intensity is seen with applied field, but this difference disappears above the ferroelectric transition between 400 and 450 K. In Figures \ref{fig:fig1}(c) and \ref{fig:fig1}(d) we show that the suppression of T1-diffuse scattering under FC conditions is direction-dependent, being absent for [100]-polarized diffuse scattering measured near (100) (Fig.\ \ref{fig:fig1}(c)) but present for [001]-polarized diffuse scattering near (001) (Fig.\ \ref{fig:fig1}(d)). These data were taken at 200 K as transverse scans across the Bragg peaks, with integration ranges of $\pm 0.5$ meV for $\hbar \omega$, and $\pm 0.05$ r.l.u.\ for the $H$ and $L$ directions for Figures \ref{fig:fig1}(c) and \ref{fig:fig1}(d), respectively. The direction dependence of the suppression of T1-diffuse scattering under [001]-field cooling is consistent with previous reports on the related PZN-x\%PT system \cite{gehring_electric-field_2004, xu_two-component_2010}.

\begin{figure}[h]
\begin{center}
\includegraphics[width=9.5cm]
{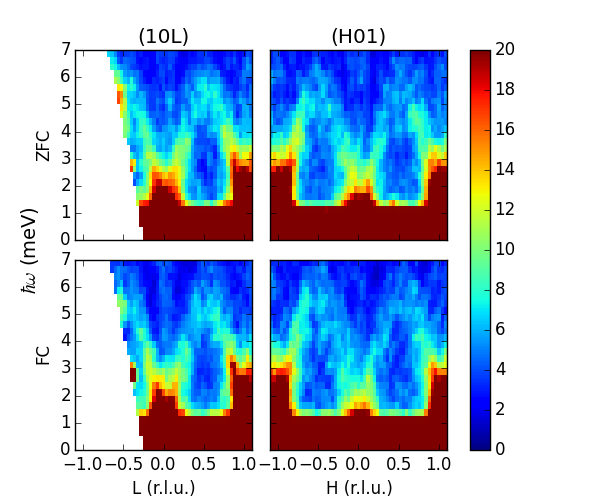}
\caption{\label{fig:TA1_slices} (Color online.) Pseudocolor plots of transverse acoustic phonons near (100) (left) and (001) (right), for ZFC (top) and FC (bottom) conditions. Neutron scattering intensity (indicated by color, in arb. units) is plotted against energy and momentum transfer. These data were taken at 400 K. White areas represent lack of data.}
\end{center}
\end{figure}

\begin{figure}[h]
\begin{center}
\includegraphics[width=9.5cm]
{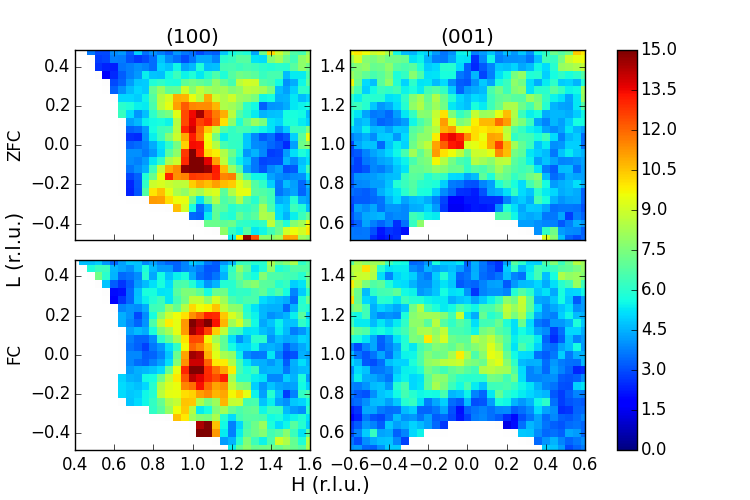}
\caption{\label{fig:TA_Eslices} (Color online.) Pseudocolor plots of transverse acoustic phonons in the $H0L$ plane near (100) (left) and (001) (right), for ZFC (top) and FC (bottom) conditions, illustrating the suppression of phonon spectral weight with field near (001) but not (100). Neutron scattering intensity (indicated by color in arb.\ units) is plotted against energy and momentum transfer. These data were taken at 400 K, within $1.5\leq \hbar \omega < 2.5$ meV. White areas represent a lack of data.}
\end{center}
\end{figure}

Fig.\ \ref{fig:TA1_slices} shows the dispersions of the TA$_1$ phonons near (100) and (001) under FZ and ZFC conditions, in which a change in intensity under field can be seen near (001) but not (100). These are pseudocolor plots of intensity vs.\ energy and momentum transfer in slices across (100) and (001) at 400 K. The intensities were integrated within $\pm 0.05$ r.l.u.\ along $H$ for (10$L$) and $L$ for ($H$01). The TA$_{1}$ phonons disperse out from the Bragg peaks and have maxima around roughly 5-6 meV. The width of the phonons with respect to energy is evident. The effect of field can be seen more clearly in Fig.\ \ref{fig:TA_Eslices}, which shows the phonon intensities near (100) and (001) in a similar pseudocolor plot, but with $\hbar \omega$ fixed at 2 meV and integrated within $\pm 0.5$ meV. There is a clear decrease in intensity with field near (001), but no clear change near (100).

\begin{figure}[h]
\begin{center}
\includegraphics[width=9.5cm]
{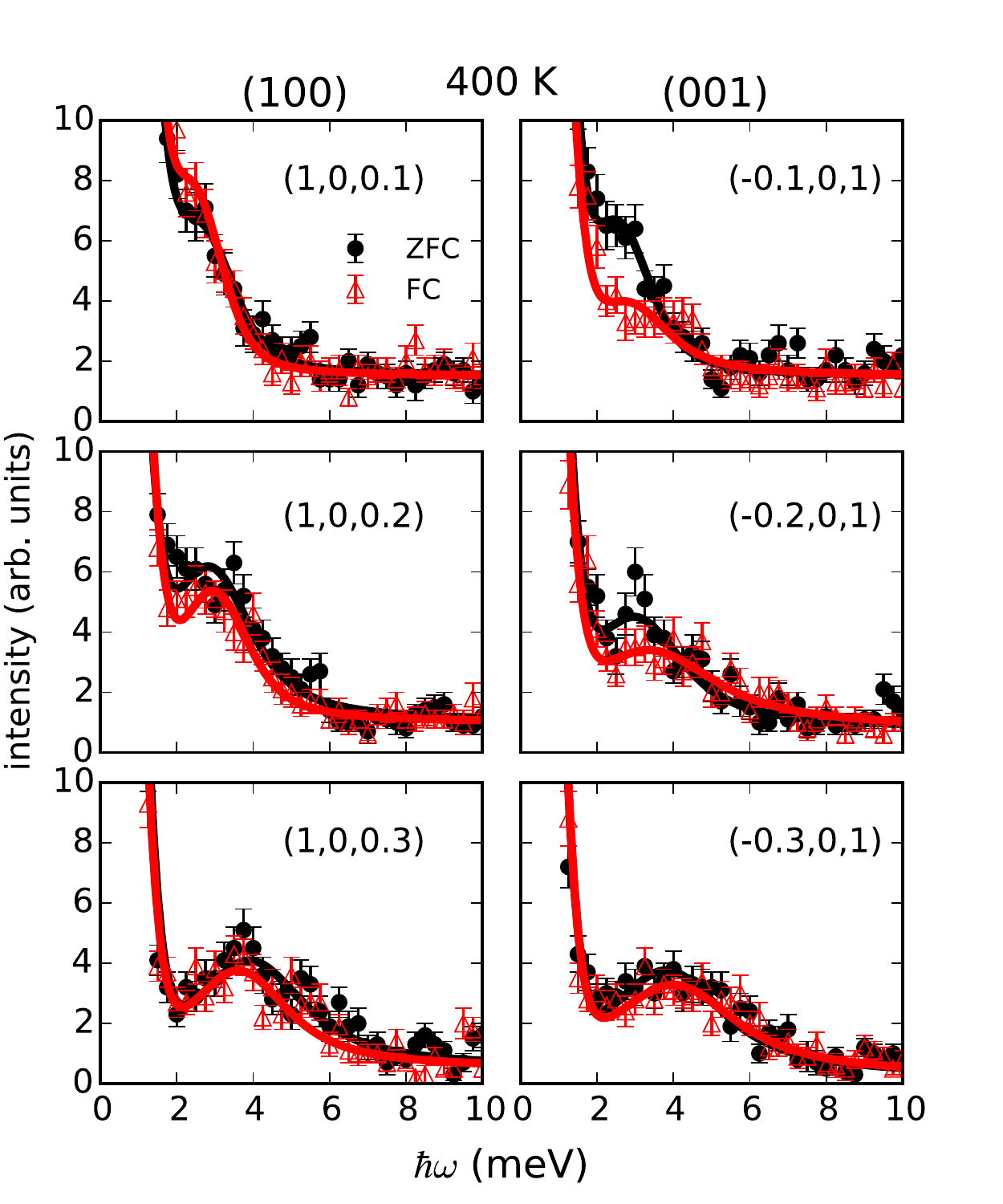}
\caption{\label{fig:e-cuts} (Color online.) Constant-$\mathbf{Q}$ cuts of transverse acoustic phonon lineshapes, with neutron scattering intensity plotted against energy transfer $\hbar \omega$. Temperatures at which data were taken are displayed at the top of each column of subplots. ZFC data represented by black circles, and FC data represented by red triangles; the plotted lines are fits through the data as described in the text.}
\end{center}
\end{figure}

For a clearer view of how the phonon dispersion is affected by field, in Fig.\ \ref{fig:e-cuts} we show constant-$\mathbf{Q}$ scans at ($-q$,0,1) and (1,0,$q$) for various $q$ values, taken at 400 K. The data were integrated over $\pm 0.05$ r.l.u.\ in both the $H$- and $L$-directions. Each data set was fitted to the sum of a Gaussian function for elastic scattering and Voigt functions for the acoustic phonons at $\pm \hbar \omega$.  We can see that there is little change near (100) for all $q$, but near (001) a clear change is seen, with both a suppression of intensity and an increase in energy transfer for $q=0.1$ (and possibly also $q=0.2$). These data suggest that the electric field effect is strongest for low $q$. We note that we have not seen a clear field effect on longitudinal acoustic (LA) phonons measured along $[100]$ near $(100)$ and along $[001]$ near $(001)$, or on TA2 phonons measured along $[\bar{1}01]$ near $(101)$, suggesting that the [001]-field primarily affects T1-phonons. This situation is similar to how T1-diffuse scattering intensities respond to [001] fields, with T1-diffuse scattering suppressed near Bragg peaks with wavevector $\mathbf{G}$ $\parallel$ [001] but unaffected for $\mathbf{G}$ $\parallel$ [100] \cite{gehring_electric-field_2004, xu_two-component_2010}, and to the lack of effect on the ($H0L$) zone T2-diffuse scattering by a [001] field \cite{wen_response_2008}.

\begin{figure}[h]
\begin{center}
\includegraphics[width=9.5cm]
{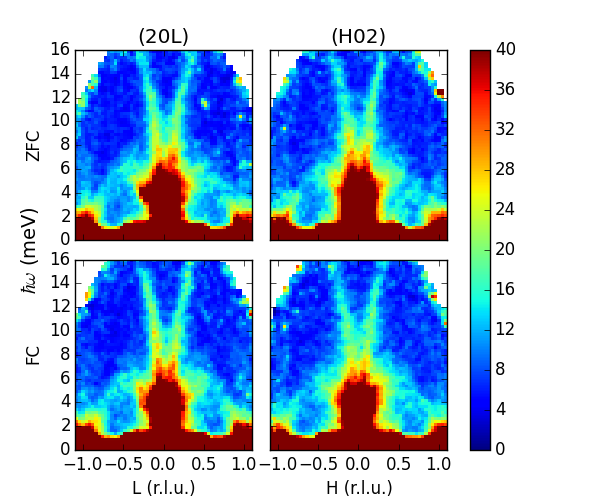}
\caption{\label{fig:slicesOptic} (Color online.) Pseudocolor plots illustrating the transverse optic modes near (200) (right) and (002) (left) for ZFC (top) and FC (bottom) conditions, with neutron scattering intensity plotted as color (in arb.\ units) as a function of energy and momentum transfer. White areas represent lack of data.}
\end{center}
\end{figure}

\begin{figure}[h]
\begin{center}
\includegraphics[width=9.5cm]
{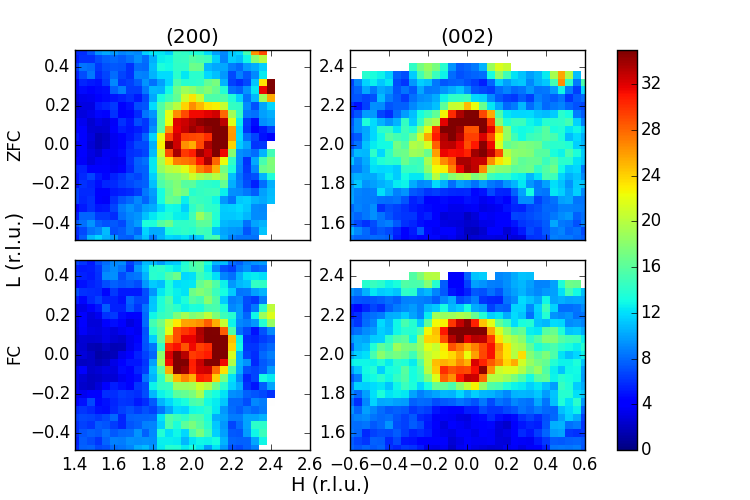}
\caption{\label{fig:TO_Eslices} (Color online.) $H$0$L$ slices at 400 K and $5.5 \leq \hbar \omega \leq 6.5$ meV, focusing on the TO phonons near (200) and (002). These are pseudocolor plots, with intensity plotted as color (in arb.\ units), and momentum transfer along $(2,0,L)$ and $(H,0,2)$ delimited on the axes. White areas represent lack of data.}
\end{center}
\end{figure}

The transverse optic modes near (200) and (002) at 400 K are shown in Fig.\ \ref{fig:slicesOptic}. (Faint spectral weight from these modes were also seen near (100) and (001) but were too weak to clearly discern.) Each panel consists of a pseudocolor plot of the scattering intensity, with energy transfer plotted on the vertical axes, and momentum transfer in the transverse direction across the Bragg peaks plotted on the horizontal axes. The data were integrated within $\pm$ r.l.u.\ along $H$ for (20$L$) and $L$ for ($H$02). The dispersion exhibits the ``waterfall effect'' seen in other PMN-x\%PT and PZN-x\%PT compositions \cite{gehring_soft_2000-1, gehring_soft_2000, gehring_dynamical_2001, gehring_soft_2001, la-orauttapong_phase_2002, koo_anomalous_2002, hlinka_origin_2003, stock_damped_2006}, where the TO$_1$ phonon softens, approaches the TA$_{1}$ mode energies, and becomes highly damped at small $q$. Unfortunately, this effect made it difficult to measure the TA$_1$ modes at small $q$ near (200) or (002), and we could not discern changes with field.

\begin{figure}[h] 
\begin{center}
\includegraphics[width=9cm]
{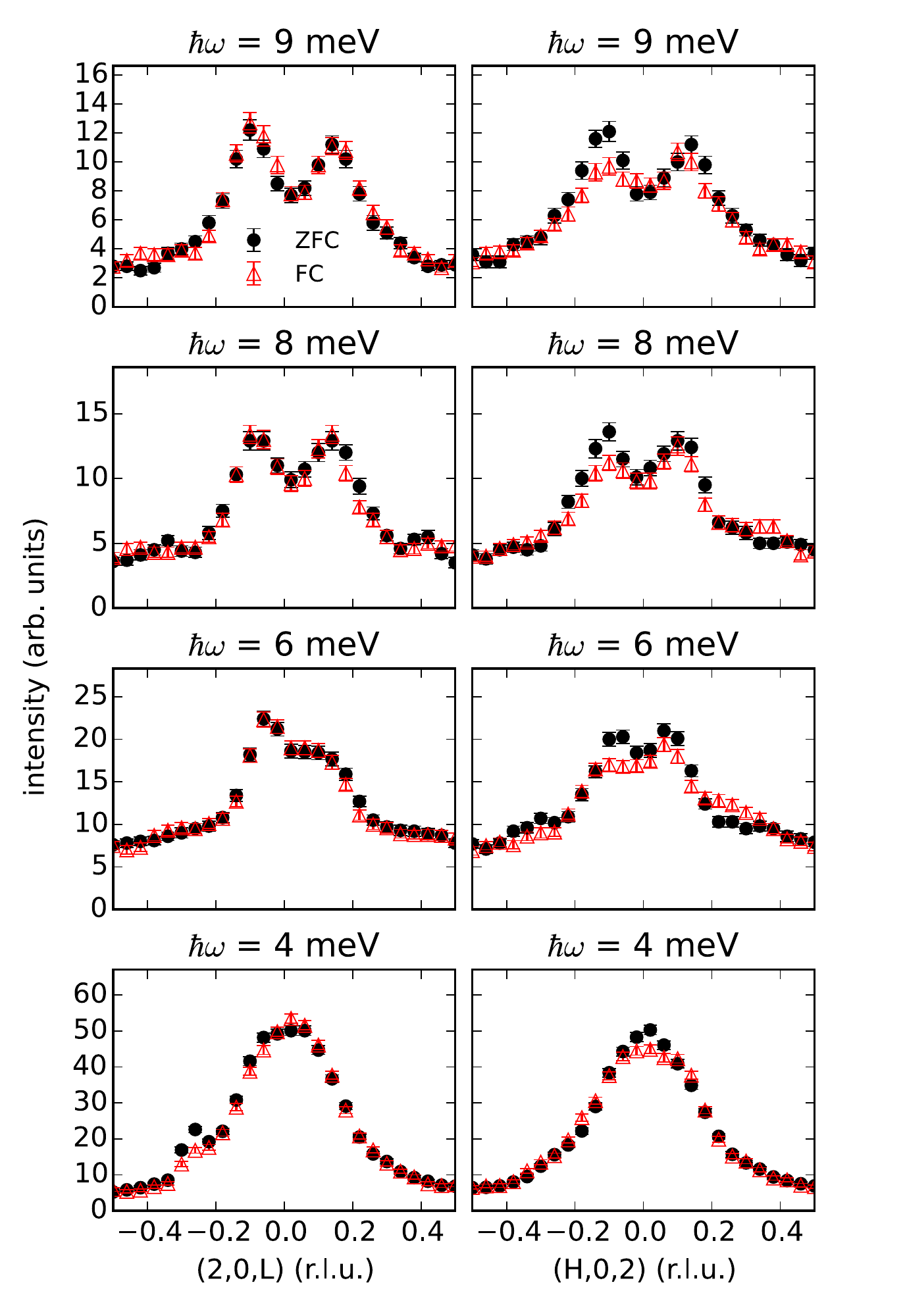}
\caption{\label{fig:cutsOptic} (Color online.) Constant-energy cuts across the transverse optic modes along (20$L$) (right) and ($H$02) (left) for $\hbar \omega =$ 4, 6, 8, and 9 meV, integrated over $1.9\leq H \leq 2.1$ for (20L) and $1.9 \leq L \leq 2.1$ for (H02), and within an energy range of $\pm 0.5$ meV. For each subplot, neutron scattering intensity is plotted against momentum transfer. All data were taken at 400 K.}
\end{center}
\end{figure}

Some suppression of spectral weight under FC conditions can be seen in Fig.\ \ref{fig:TO_Eslices}, which shows pseudocolor plots of constant-energy slices at $\hbar \omega = 6$ meV and $400$ K, with intensities integrated within $\pm 0.5$ meV. Specifically, a slight decrease can be seen near (002) with field-cooling, but no change is clear near (200). For a clearer view of the spectral weight suppression, Fig.\ \ref{fig:cutsOptic} shows constant-energy cuts made along the transverse directions across the (200) and (002) Bragg peaks. In each panel, scattering intensity is plotted against momentum transfer for data taken under ZFC and FC conditions. Intensity was integrated within $\pm 0.5$ meV for energy transfer and $\pm 0.1$ r.l.u.\ for momentum transfer in the $H$ or $L$ direction transverse to the direction of the scan. We see that there is a consistent suppression of spectral weight near (002) but not near (200). This suppression can be seen from 4 to 9 meV; we note that the difference disappears outside of this range. As for TO$_{1}$ phonon energy, it is difficult to observe changes in the TO$_{1}$ dispersions due to their steepness and to the phonons becoming highly damped at small $q$. For comparison, we note that in conventional ferroelectrics there have been examples of optic modes being affected by field \cite{worlock_electric_1967, watanabe_brillouin_1998}, and the effect is only predicted to be large for soft modes close to zero energy 
in the vicinity of a structural phase transition \cite{boccara_electric_1968}.

\begin{figure}[h]
\begin{center}
\includegraphics[width=9cm]
{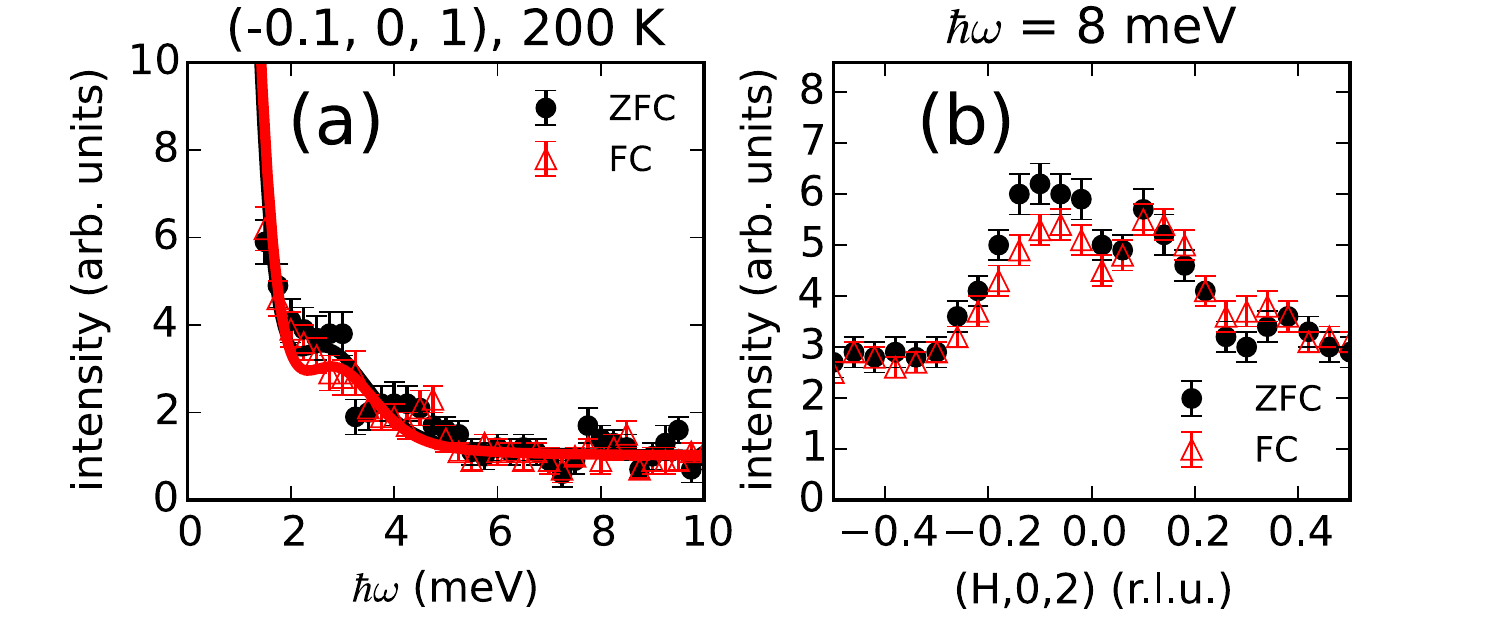}
\caption{\label{fig:figs200K} (Color online.) Neutron scattering scans at 200 K showing less change between ZFC and FC conditions near (001) and (002) than in the 400 K data in Figures \ref{fig:e-cuts} and \ref{fig:cutsOptic}. (a) Constant-$\mathbf{Q}$ scan at $(-0.1, 0, 1)$, with intensity plotted against energy transfer. (b) Constant-$\hbar \omega$ scan at $\hbar \omega = 8$ meV, with intensity plotted against momentum transfer.}
\end{center}
\end{figure}

To illustrate the effect of temperature, in Fig.\ \ref{fig:figs200K} we show representive data of the TA$_1$ and TO$_1$ modes at 200 K to contrast with the 400 K data in Figures \ref{fig:e-cuts} and \ref{fig:cutsOptic}, respectively. In Fig.\ \ref{fig:figs200K}(a), we show a constant-$\mathbf{Q}$ scan showing the TA$_1$ mode at $(-0.1, 0, 1)$, with intensity vs.\ $\hbar \omega$ plotted, and intensity integrated within $\pm 0.05$ r.l.u.\ in the $H$ and $L$ directions. In Fig.\ \ref{fig:figs200K}(b), we show a constant-$\hbar \omega$ scan showing the TO$_1$ mode, integrated within $7.5 \leq \hbar \omega \leq 8.5$ meV and $1.9 \leq L \leq 2.1$ r.l.u. In both plots, we see a similar suppression of intensity near (001) and (002). We also saw a similar lack of change near (200) (not plotted), but near (100) a spurious feature prevented us from determining if there was a change in TA$_{1}$ phonon spectral weight. From these data, we can see that the field effect at 200~K seems to be less pronounced than that at 400~K, at least for the acoustic phonon mode.

\section{Discussion}
The electric field effects observed in our measurements can be summarized as (i) there is no [001] field effect on the longitudinal acoustic (LA) modes, or on the transverse acoustic modes propagating along the $\langle 110 \rangle$ (TA$_2$) directions; (ii) for TA$_{1}$ phonons polarized along $\langle 001 \rangle$, we observed, after field-cooling, a reduction of intensity and increase of phonon energy near Bragg peak wavevectors $\mathbf{G}$ $\parallel$ [001], but no field effect was observed for TA$_1$ modes near $\mathbf{G}$ $\parallel$ [100]. For diffuse scattering, a similar pattern in response to field-cooling along [001] has been seen, with intensity suppression for $\mathbf{G}$ $\parallel$ [001], but not for $\mathbf{G}$ $\parallel$ [100] \cite{gehring_electric-field_2004, xu_two-component_2010}. A much smaller effect on low energy TO modes is also observed following the same rule, i.e. a reduction of intensity near the bottom of the TO mode measured for $\mathbf{G} \parallel$ [001], but no effect for $\mathbf{G} \parallel$ [100]. These results imply that a coupling exists between the diffuse scattering along $\langle 001 \rangle$ (the T1-diffuse) and the TA$_1$ and/or TO$_1$ phonon modes along the same directions.

Previous work has shown a strong coupling between the diffuse scattering along the $\langle 110 \rangle$ directions (the T2-diffuse) and the TA$_2$ phonon modes in these lead-based relaxor materials~\cite{xu_phase_2008}. The diffuse-phonon coupling along the $\langle 001 \rangle$ (T1) and $\langle 110 \rangle$ (T2) directions share some common features. For example, when the diffuse scattering is suppressed by the external field, we always see a hardening of the corresponding TA phonon mode. Evidentally, interaction between the PNR and the phonons tends to drive the phonons softer, for both the T1 and T2 modes. This tendency suggests that short-range polar orders are likely related to lattice instabilities in these relaxor compounds.

On the other hand, the T1 and T2 modes differ in many aspects too: 

(i) For the case of the T2 modes, the field effect on the PNR (and thus on the T2-diffuse scattering) is indirect. An external field along [111] helps establish a single domain ferroelectric phase with [111] polarization. The change in the population of each domain induces a redistribution of PNR with different polarizations, resulting in the redistribution of T2-diffuse scattering intensities in reciprocal space. For example, under a [111] field there is an enhancement of the T2-diffuse scattering near the (220) Bragg peak, but a reduction of the T2-diffuse scattering near the ($\bar{2}$20) Bragg peak. The TA$_{2}$ phonon near (220) softens, while the phonon near ($\bar{2}$20) hardens. On the other hand, in zero field, all $\langle 111 \rangle$ domains are present to an equal degree. Measurements in zero field of the TA$_{2}$ phonons near (220) or ($\bar{2}20$) average over both hardened and softened phonons, and the role of the [111] field is merely to obtain results for a single domain. 

However, the [001] field effects on the T1-diffuse scattering and T1-phonons are not domain related. This can be demonstrated by comparing the ZFC and FC results. Under FC conditions, one only sees changes (relative to the ZFC results) for $\mathbf{G}$ $\parallel$ [001], but no change for $\mathbf{G}$ $\parallel$ [100]. If these changes were due to domain effects, one would have expected the ZFC results to lie in between (in intensity, energy, etc.) the FC results measured for $\mathbf{G}$ $\parallel$ [100] and [001]. This is not the case. Therefore, the effect of the [001] field is more intrinsic, with the field directly affecting the short-range order and consequently the related phonon modes. 

(ii) While there is no evidence of diffuse-TO coupling for the T2 modes \cite{xu_phase_2008}, in our sample there appears to be a weak diffuse-TO coupling for $\mathbf{G}$ $\parallel$ $\langle 001 \rangle$ (TO$_{1}$ modes). 

(iii) The diffuse-phonon coupling for the T2 modes is strong throughout the entire Brillouin zone, but the coupling for the T1 modes is only present for a small range of $q$ values ($\sim$0.1 to 0.2 r.l.u.)~away from $\mathbf{G}$. This anisotropy of the diffuse-TA coupling revealed by our electric field measurements is also consistent with previous reports~\cite{stock_evidence_2012}.

In order to understand these results, we consider the origin of the diffuse scattering and short-range orders in relaxors. The local random electric field generated by the B-site cations are believed to play important roles in Pb-based relaxor systems~\cite{vugmeister_dipole_1990, westphal_diffuse_1992, stock_universal_2004, tinte_origin_2006, cowley_relaxing_2011, pirc_spherical_1999}. A direct link between the diffuse scattering in these relaxors and the random field has been demonstrated by comparing two isostructures with and without random B-site valences~\cite{phelan_role_2014}. The random field in the system prevents long-range order from developing and induces short-range orders that also grow with cooling. The diffuse scattering intensities from these short-range orders are not entirely static, and have a strong dynamic component~\cite{xu_two-component_2010, xu_freezing_2012, stock_interplay_2010}. The existence of these dynamic/quasi-elastic components has been explained by theoretical work~\cite{vugmeister_polarization_2006, pirc_dynamics_2001}, and is essential for the coupling between the diffuse scattering and phonons. Indeed one can see that in our work, the coupling between the T1-diffuse scattering and the TA$_1$ phonon is weaker at 200~K than 400~K, where the dynamic component has also been shown to decrease with cooling~\cite{xu_freezing_2012, stock_interplay_2010} below T$_C$. The weak coupling to the TO$_1$ mode, can be understood based on knowing that the short-range orders consist of a combination of acoustic (strain) and optic (polar) types of atomic shifts~\cite{xu_coexistence_2006, xu_two-component_2010}. The polar component of the short-range order can couple to the TO mode. The coupling will likely diminish quickly when the TO phonon energy increases and moves further away from the quasi-elastic component of the diffuse scattering, as is the case in our measurements.

The anisotropy of the diffuse-phonon coupling between the T1 and T2 directions is more intriguing.  The T2-diffuse is significantly stronger than the T1-diffuse, extends to a larger $q$-range in reciprocal space, and interacts with TA phonons along almost the entire branch. Overall, we could consider a picture where PNRs are actually the ``core'' of the short-range correlations in the system, and contribute to the broader T2-diffuse; while the polar/strain field surrounding the core can extend to much larger range and contribute to the narrower T1-diffuse. The atomic shifts in the PNRs would be significantly larger than those in the surrounding region, leading to a much stronger T2-diffuse scattering than the weaker T1-diffuse. The core of the short-range order, i.e. the PNRs,  results from the local strong random field and cannot be directly suppressed by an external field~\cite{xu_coexistence_2006, xu_electric-field-induced_2006, xu_phase_2008}. However, the weaker polar/strain field around the core is less robust and can be partially modified by external field, showing the intrinsic field effect on the T1-diffuse and its coupling to TA$_1$ and TO$_1$ phonon modes discussed in this paper. 

An analog to this situation has recently been considered~\cite{okamoto_experimental_2015, proctor_effect_2015} where strong but dilute random fields are inserted into a system with a weak continous random field. Theoretical work involving a magnetic system with a random field \cite{proctor_effect_2015} suggested that a large correlation length or even a weak long-range order could be achieved. If we map the PNRs to the strong random field in the magnetic system, the large spin correlation length proposed by the theoretical work can be related to the weak polar/strain field surrounding the PNRs which gives the T1-diffuse. Although not an exact analog, this picture does provide a crude description of the origin of the two types of diffuse scattering. For a better understanding of the source of these diffuse scattering components and their coupling to lattice dynamics, more detailed experimental work is required. Though numerous models have been proposed by various groups trying to describe the diffuse scattering and short-range orders in these relaxor systems~\cite{welberry_single-crystal_2005, xu_three-dimensional_2004, welberry_chemical_2006, pasciak_interpretation_2007, ganesh_origin_2010, vakhrushev_diffuse_2005, burkovsky_inelastic_2010, bosak_diffuse_2012, cervellino_diffuse_2011}, our results simply suggest that there is a clear anisotropy for diffuse scattering, their field dependence, and their coupling to the related phonon modes measured along  $\langle 100 \rangle$ and $\langle 101 \rangle$ directions, and do not favor any particular model.

\begin{table}
\caption{\label{table:elasticConstants} $C_{44}$ and $(C_{11} - C_{12})/2$ elastic constant data from neutron scattering experiments in units of 10$^{11}$N/m$^{2}$. PMN-32\%PT values calculated from data taken at 400 K; all other values calculated from data at 300 K.}
\begin{ruledtabular}
\begin{tabular}{lll}
Material 							& 	$C_{44}$ 		& 	$(C_{11} - C_{12})/2$   \\
\hline \\
PMN	\cite{stock_evidence_2012}		&	0.53(3)		&	0.48(6)				\\
PMN-32\%PT						&	0.56(5)		&	0.23(4)				\\
PZN-4.5\%PT \cite{xu_phase_2008}	&				&	0.26(4)				\\
PbTiO$_{3}$ \cite{tomeno_lattice_2006}	&	0.72(2)		&	0.63(1)				\\
\end{tabular}
\end{ruledtabular}
\end{table}

We can compare elastic constants derived from the TA$_{1}$ and TA$_{2}$ phonon energies in our data with values reported for similar materials to get insight into the tendency for lattice instability in the T1 and T2 directions. Our values and those for related compounds in the literature are displayed in Table \ref{table:elasticConstants}. From our data, we obtained the elastic constant quantities $C_{44} = 0.56(5)$ and $(C_{11}-C_{12})/2=0.23(4)$ in units of $10^{11}$~N/m$^2$ based on the TA$_1$ and TA$_2$ phonons measured near (001) and (101) at $T=400$ K. The value of $C_{44}$ is slightly larger than in PMN but smaller than in PbTiO$_{3}$~\cite{stock_evidence_2012}. $(C_{11}-C_{12})/2=0.23(4)$ shows a bigger change, being significantly reduced from its value in PMN, suggesting an increased lattice instability when the system approaches the morphotropic phase boundary. In fact, a similar value of $(C_{11}-C_{12})/2=0.26(4)$ can be obtained from phonon data on PZN-4.5\%PT~\cite{xu_phase_2008}. 

The diffuse-phonon coupling discussed in this paper can also affect how one determines the elastic constants. We calculated the phonon velocities from our FC data rather than our ZFC data since we believe the FC velocities more closely resemble the velocities expected for $q \rightarrow 0$. First, the diffuse-phonon coupling is expected to diminish as $q \rightarrow 0$ \cite{axe_anomalous_1970}. Second, since the diffuse-phonon coupling tends to reduce the phonon energy (as discussed in this paper and Refs.~\cite{xu_phase_2008, stock_evidence_2012}), at non-zero $q$ the elastic constants should be calculated from data taken after this coupling effect is removed, such as after diffuse scattering is suppressed by field. For ZFC conditions, the value for $C_{44}$ can be about 15\% smaller than in FC conditions. This difference is in fact an artifact of calculating the  phonon velocity using (reduced) phonon energies at non-zero $q$ (we used data taken at $q=0.1$ and $q=0.1 \times \sqrt{2}$ r.l.u.\ for the calculation in the T1 and T2 directions, respectively). If one could obtain the phonon velocity using smaller $q$ values near $q=0$ where the diffuse-phonon coupling diminishes, the difference between ZFC and FC data should become negligible.

\section{Conclusions}
We have observed, when comparing field-cooling and zero-field-cooling conditions for different Brillouin zones with a field along [001], a change in the lattice dynamics of PMN-32\%PT that correlates with changes in diffuse scattering. Specifically, under field cooling we see a reduction of intensity and an increase of phonon energy for the TA$_{1}$ mode measured near (001) and propagating along [100] ($\langle 001 \rangle$-polarized), but no change for TA$_{1}$ phonons near (100) and propagating along [001] ($\langle 100 \rangle$-polarized). This field effect is only clearly seen for wavevectors around 0.1 to 0.2 r.l.u.\ away from the Bragg peak. Meanwhile, the T1-diffuse scattering near (001) is suppressed under field-cooling, but is unaffected near (100). A similar effect is seen for the TO$_{1}$ mode, which is slightly suppressed near (002) from 4 to 9 meV, but unaffected near (200). No clear field effect has been seen for the longitudinal modes near (001) or (100), or for the TA$_{2}$ mode near (101). The similarities in the effect of field on the T1-diffuse scattering near (001) and (100), the TA$_{1}$ phonons near (001) and (100), and the TO$_{1}$ phonons near (002) and (200) suggest the presence of diffuse-TA and diffuse-TO mode coupling which resembles the mode coupling observed in the T2 directions in related relaxor materials.

\section*{Acknowledgements}
This research at Oak Ridge National Laboratory's Spallation Neutron Source was sponsored by the Scientific User Facilities Division, Office of Basic Energy Sciences, U.S. Department of Energy. J.A.S.\ and G.Y.X.\ acknowledge support by Office of Basic Energy Sciences, U.S. Department of Energy under contract No.\ DE-SC00112704. Z.J.X.\ and R.J.B.\ are also supported by the Office of Basic Energy Sciences, U.S.\ Department of Energy through Contract No.\ DE-AC02-05CH11231. C.S.\ acknowledges the support of the Carnegie Trust for the Universities of Scotland and the Royal Society. The identification of any commercial product or trade name does not imply endorsement or recommendation by the National Institute of Standards and Technology.

\end{document}